\documentclass[fleqn,twoside,twocolumn,nofootinbib,showkeys]{revtex4} 
\usepackage[sec,nocpr]{ujp} 
\begin{document}
\title[Bias of the Hubble constant value caused by errors in galactic distance indicators]
{BIAS OF THE HUBBLE CONSTANT VALUE CAUSED BY ERRORS IN GALACTIC DISTANCE INDICATORS}%
\author{S.L.~Parnovsky}
\affiliation{Taras Shevchenko National University of Kyiv}
\address{Observatorna str., 3, 04058, Kyiv, Ukraine}
\email{parnovsky@knu.ua}

\udk{524.8} \pacs{98.80.Es, 02.70.Rr} \razd{\seci}

\autorcol{S.L.~Parnovsky}

\setcounter{page}{1}%

\begin{abstract}
The bias in the determination of the Hubble parameter and the Hubble constant in the modern Universe 
is discussed. It could appear due to statistical processing of data on galaxies redshifts and estimated distances   
based on some statistical relations with limited accuracy. This causes a number 
of effects leading to either underestimation or overestimation of the Hubble parameter when using any methods of statistical processing, 
primarily the least squares method (LSM). The value of the Hubble constant is underestimated when processing a whole sample; when the 
sample is constrained by distance, especially when constrained from above, it is significantly overestimated due to data selection. The 
bias significantly exceeds the values of the error the Hubble constant calculated by the LSM formulae.

These effects are demonstrated both analytically and using Monte Carlo simulations, which introduce deviations in both velocities and 
estimated distances to the original dataset described by the Hubble law. The characteristics of the deviations are 
similar to real observations. Errors in estimated distances are up to
$ 20\%$. They lead to the fact that when processing the same mock sample using LSM, it is possible to obtain 
an estimate of the Hubble constant from $96 \%$ of the true value when processing the entire sample to $110\%$ when processing the subsample with 
distances limited from above.

The impact of these effects can lead to a bias in the Hubble constant obtained from real data and an overestimation of the accuracy 
of determining this value. This may call into question the accuracy of 
determining the Hubble constant and significantly reduce the tension between the values obtained from the observations in 
the early and modern Universe, which were actively discussed during the last year.
\end{abstract}

\keywords{cosmology, cosmological parameters, Hubble constant tension, methods: statistical}

\maketitle

\section{Introduction}\label{s1}
The basis of modern cosmology is a homogeneous isotropic model, in which all points in the Universe and all directions are
equivalent. Two details need clarification. It is possible that the Universe was anisotropic at the Big Bang time, but it became 
almost isotropic in the 
era of inflationary expansion in tiny fractions of a second and has been almost isotropic ever since. The temperature of the cosmic 
microwave background radiation is almost the same in all directions. This indicates a high degree of homogeneity and isotropy of the 
Universe during the epoch of recombination. However, as the Universe expanded, the fluctuations grew, e.g. of the density of the matter 
filling it. This led to the formation of a large-scale structure and the appearance of superclusters, voids, clusters, galaxies and 
stars. At present, we can talk about the homogeneity of the Universe only on very large spatial scales.

The rate of expansion of the Universe in such models is characterized by the time-dependent Hubble parameter $H$. It is defined as 
$H=\dot{a}/a$, where $a(t)$ is the scale factor and dot means the derivative with respect to cosmological time $t$.
Its current value is called the Hubble constant and is denoted by $H_0$. Astronomers use the associated dimensionless quantity $h$, 
defined as $H_0 = h\cdot 100$ km s$^{-1}$ Mpc$^{-1}$. With its help, Hubble velocities are easily converted into distances measured in 
Mpc $h^{-1}$. The modern estimate gives the value $h\approx 0.7$.

Most measurements of the Hubble parameter deal with
distances, which are small by cosmological standards. They have small redshifts 
and relate to the late Universe. However, a few measurements relate to the early Universe, e.g. to 
the recombination era (redshift $z\approx 1100$). 
First of all, these are CMB data from Planck satellite \cite{pl} and data from Dark Energy Survey Year 1 clustering combined with 
data on weak lensing, baryon acoustic oscillations and Big Bang nucleosynthesis \cite{abb}. The data on primary nucleosynthesis deal 
with a process that took place in the first minutes of the existence of the Universe.

Estimations of the Hubble constant obtained by different methods are given in \cite{r}. They are obtained from different 
observational data and based on different physical effects and its common constraints. The values given there, 
namely the estimation $H_0=67.4\pm0.5$ km s$^{-1}$ Mpc$^{-1}$ in the recombination era and 
$H_0=73.3\pm0.8$ km s$^{-1}$ Mpc$^{-1}$ in modern era are differ by 8\%. The corresponding difference is at the level of $4\sigma-6\sigma$, which, according 
to \cite{r}, should be classified as something from a discrepancy or a problem to a crisis. 

Numerous articles have tried to explain this difference in a variety of ways, including the introduction of new physical interactions 
in the early Universe. Naturally, the possible connection between the Hubble tension and the violation of the isotropy or homogeneity 
of the Universe was also considered.
The review \cite{2021arXiv210301183D} contains links to 882 papers related to this issue. Some attempts 
to explain Hubble tension are discussed in \cite{Efstathiou, Schoneberg, bv}. According to \cite{Efstathiou1} a 
systematic bias of $\sim$ 0.1 -- 0.15 mag in the intercept of the Cepheid period-luminosity relations of SH0ES galaxies could resolves 
the Hubble tension. Note that in the paper \cite{Freedman} the value of $H_0 = 69.6 \pm 0.8$ ($\pm 1.1\%$ stat) $\pm 1.7$ ($\pm 2.4\%$ 
sys) km s$^{-1}$ Mpc$^{-1}$ was find for the late Universe using the direct revised measurement of the tip of the red giant branch of 
the Large Magellanic Cloud. It agrees well with the ones for the early Universe.

I want to emphasize that the purpose of this work is neither to explain the differences in the estimates of $H_0$ for the early and late 
Universe due to the impact of some unaccounted factor nor to criticize any results obtained earlier. I want to demonstrate the pitfalls 
that always exist in data processing. 

For this I am using the simplest example. The Monte Carlo method makes it easy to study the 
accuracy and precision of the estimates obtained; the use of the simplest least squares method with a single determinable parameter 
makes it possible not to consider the details of the complex analysis used in the processing of real data. I am not discussing Fisher 
matrices, possible abnormal distribution of deviations, discarding outliers, and other important processing details. My goal is to 
show that even in this simple case, the use of standard processing methods, but with different processing details, for example, the 
cut-off boundaries of the used subsamples, can give results that are quite different from each other. The formal application of 
statistical criteria could lead to the conclusion that there are significant differences in results of processing the same initial 
data set using the same method and the same procedure for its application.
 
I consider the issues of statistical processing of data on redshifts and estimating distances of galaxies in the modern 
Universe. I demonstrate that the differences in $H_0$ 
estimations could be explained by quite trivial reasons, such as the usage of statistical dependencies when estimating distances 
to galaxies.

To estimate distances to galaxies independently from redshifts astronomers have to use some distance 
indicators such as distance from Cepheid variables, the Tully-Fisher relation for spiral galaxies, the $D_n-\sigma$ or fundamental 
plane relations for elliptical galaxies, surface brightness fluctuations, brightest cluster galaxies, or tip of the red-giant branch. 
All of them are based of different statistical relations and provide distances with different accuracy. 

The most precise of these 
statistical dependencies can only be used to estimate the distances to nearby galaxies, whose radial velocities exceed only slightly the 
characteristic speed of collective non-Hubble motions. For example, 
in the paper \cite{2021ApJ...908L...6R} luminosities of 75 Cepheids from the Milky Way were measured with errors of 1\%. These 
data were used for calibration the distances. The estimate $H_0 = 73.0 \pm 1.4$ km s$^{-1}$Mpc$^{-1}$, which differs by 
4.2 $\sigma$ from the Planck results, was obtained using distance indicators based on these new data.

Tully-Fisher relation was generally considered to give the best distances to galaxies in the range required to define $H_0$ for some 20 
years after its inception in \cite{TF}. It used a correlation for spiral galaxies between their luminosity and how fast they are 
rotating. The latter can be determined from the galaxy emission line widths. R.B. Tully wrote in his Scholarpedia article about the 
Tully-Fisher relation that at optimal passbands between 600 nm and 800 nm, its scatter is $\sim$0.35 magnitudes, equivalent to 17\% 
uncertainty in distance. For other ranges or when using the dependence of the linear diameter on the emission line width, this 
uncertainty is bigger. This value is one of the parameters I use in Monte Carlo simulations. In modeling, I consider relative 
errors in determining the distance to individual galaxies (without using their redshift data) up to 20-30\% based on the accuracy 
of the Tully-Fisher method. However, the accuracy of determining the Hubble constant is significantly higher due to the fact 
that statistical processing of a large array of data on galaxies is used.

Note that when estimating the distance from supernova Ia explosions, astronomers deal directly with the light curve and the spectrum 
from which the redshift can be determined. This was enough to discover the accelerated expansion of the Universe. Explosions can be 
considered the ``standard candle'' so beloved by cosmologists. 

However, when determining the Hubble constant, the distance scale to them 
must be calibrated. This is done on the basis of data on the distances to galaxies in which a supernova explosion occurs, obtained
from some of the criteria described above and retain all the systematic 
biases inherent in them. An overview of the methods used to determine the Hubble constant and construct a ladder of distances to distant 
extragalactic objects is given, in particular, in the review \cite{2015LRR....18....2J}.

Whichever distance estimation method is used, one gets a set of estimated distances $R_i$ instead of true distances $r_i$ to galaxies.
The difference between $R_i$ and $r_i$ are errors in estimating the distance to each galaxy we are talking about. They are not so small, 
since they accumulate all the errors inherent in all distance determination techniques used to estimate and calibrate the statistical 
dependence on all levels of the cosmic distance ladder. It is important that they increase with distance. 
Their impact leads to the bias in the value of the Hubble constant 
and its error could be significantly underestimated.

There are several sources of error in the determination of the Hubble parameter and the Hubble constant derived 
from it. One of them is the well-known bias, which arises because of errors in the argument of the function. It occurs during any 
statistical processing, including the least squares method (LSM) and the maximum likelihood estimation (MLE) for 
statistical data processing. It is discussed 
in the Section \ref{s2}. The second one arises due to data selection effects when processing data from subsamples limited by distance obtain 
by indicators $R$. They are considered in the Section \ref{s3}. 
I demonstrate their impact in subsections \ref{ss23a}, \ref{ss24a} using the Monte Carlo method for mathematical modelling. 

I also consider some potential sources of error that appeared to have no significant effect on the Hubble constant value. 
For example, in subsection \ref{ss25} the influence of collective motions of galaxies aka cosmic streams is considered.
Analysis show that it is not a source of errors when getting the $H_0$ value.

I considers and analyze some of the mathematical features of the used transformations in the subsection \ref{ss21a} in the Appendix. 
In the subsection \ref{ss22a} formulae for the  are derived.

I sequentially consider several effects that affect the values of $H$ and $H_0$ obtained by processing 
observational data. I start with the simplest model, gradually adding additional factors related to the model used, sample 
completeness and processing details.

\section{Underestimation of the Hubble constant caused by errors in galaxy distances}\label{s2}

\subsection{The simplest model of motion of galaxies and errors in distance and velocities measurements}\label{ss21}

Consider a sample of galaxies with distances $r_i$ and radial velocities $v_i$. At small redshifts $z\ll 1$ they are related 
by the Hubble law aka the Hubble-Lema\^{\i}tre law
\begin{equation}\label{e1}
v_i=Hr_i,
\end{equation}   
where $H$ is the Hubble parameter. Therefore, one can exactly determine its value from any sample of $r_i$ and $v_i$. But the world 
is not so ideal and instead of this sample we have to process slightly different data.
 
It is possible to determine with great precision the actual measured radial velocity of each individual galaxy $V_i$ from its redshift. 
It is the sum of the velocities of its Hubble motion $v_i$ and the radial component of the peculiar motion of this galaxy. The last 
one is the sum of radial component of the collective non-Hubble motion of galaxies or galaxy flows and some random motion of this individual 
galaxy.

We start from the simplest model without collective non-Hubble motion. It does not consider significant variations in the 
density of matter existence in different regions of space. It is the reason for the appearance of non-Hubble flows of galaxies. When processing 
real data, we take into account these collective non-Hubble motions either by using their multipole expansion \cite{2013Ap&SS.343..747P},
or by directly simulating the influence of attractors and voids. In Section \ref{ss25} I consider the influence of collective motion in 
the bulk motion approximation. However, let me remind you that my main goal is to show the pitfalls associated with statistical 
processing. And this is more clearly manifested when using the simplest models. So, I assume 
\begin{equation}\label{e2}
V_i=v_i+\delta Vs_i,
\end{equation}   
where $s_i$ is a random variable with normal distribution, zero mean and unit variance and $\delta V$ is the characteristic value of 
the random component of galaxy radial velocity. 

We deal with a set of estimated distances to galaxies $R_i$ which are different from true ones $r_i$. One can assume that
\begin{equation}\label{e3}
R_i-r_i=ar_ip_i,\; R_i=r_i(1+a p_i),
\end{equation}   
where $p_i$ is also a random variable with normal distribution, zero mean and unit variance and $a$ is the characteristic value of 
the relative error of the distance indicator we use.

Instead of (\ref{e3}) one can use 
\begin{equation}\label{e3a}
R_i-r_i=aR_ip_i,\; R_i=r_i(1-a p_i)^{-1}.
\end{equation}   
It is close to (\ref{e3}) for small errors in the distance estimation. In this case the error is a certain percent of $R_i$. 
I discuss the differences between these two types of noise (\ref{e3}) and (\ref{e3a}) in the Sect. \ref{ss21a} and show that they are 
very different from a mathematical point of view. In addition to analytical consideration I use equations (\ref{e2}) and (\ref{e3}) or 
(\ref{e3a}) to prepare a lot of mock samples for the Monte Carlo simulations. More details 
I discuss in the Section \ref{ss23}.

What type of relation can be expected from real astronomical observations? Statistical dependencies make it possible to estimate a 
certain parameter of the galaxy, which does not depend on the distance to it. Usually, this is its absolute luminosity $L$. For galaxies 
from the FGC and RFGC catalogues \cite{FGC,RFGC} this is their linear diameter $D$. The errors of these values, if they are distributed 
over a gaussian, give the relation (\ref{e3}). To determine the distance, we need to use the flux from the galaxy, i.e. its apparent 
luminosity, or its angular diameter. Errors in these values, with their normal distribution, provide the relation (\ref{e3a}). 
In general, we get a certain combination of relations (\ref{e3}) and (\ref{e3a}).

I do not use this more general option. Relations (\ref{e3}) and (\ref{e3a}) are quite enough for generating mock samples and derivation of 
formulae. After all, the purpose of our mathematical modelling is 
to demonstrate the effects and their rough estimate, and not an attempt to obtain particularly accurate estimates, taking into account 
all the nuances of real samples and methods of their processing.

\subsection{Least squares data processing}\label{ss22}

I use the standard least squares method (LSM) formulae when processing mock samples data. LSM provides the optimal proportional 
relationship $V=AR$ between the $V_i$ and $R_i$ for the sample. If the statistical weights of the data points are the same, then the slope 
coefficient can be found by the formula 
\begin{equation}\label{e40}
A=\frac{\sum_{i=1}^{N} V_i R_i}{\sum_{i=1}^{N} R_i^2}.
\end{equation}   

However, its slope coefficient would be equal to $A=kH$ instead of 
$H$ in (\ref{e1}). The factor $k$ characterizes the deviation 
of the values of the Hubble parameter and the Hubble constant obtained from true ones.
 
All odd plain central moments for a Gaussian distribution are zeroed. The mean values of $p_i^n$ are equal to $(n-1)!!$ if $n$ is even. 
Here $(n-1)!!$ denotes the double factorial, that is, the product of all odd numbers from $n-1$ to 1.
The theoretical mean value of $k$ for the noise (\ref{e2}) and (\ref{e3}) is easy to estimate as follows
\begin{equation}\label{e4}
\begin{array}{l}
k=\frac{\langle A\rangle}{H}= \bigg \langle \frac{\sum (Hr_i+\delta Vs_i) r_i(1+a p_i))}{H\sum r_i^2(1+a p_i)^2} \bigg \rangle\\ 
\approx \frac{\langle \sum r_i^2 (1+a p_i)\rangle}{\langle \sum r_i^2(1+a p_i)^2\rangle} =\frac{1}{1+a^2}.
\end{array}
\end{equation}   
Here the angle brackets mean averaging over the quantities $s_i$ and $p_i$. The average value of the numerator is $\sum Hr_i^2$. 
The average value of the denominator is $(1+a^2)\sum Hr_i^2$. The average ratio slightly differs from the ratio of these average 
values due to the correlation between the terms with $p_i$ in the numerator and denominator, but for small $a$ this can be neglected.

The equation (\ref{e40}) provides the ratio of series for the noise (\ref{e3a}). 
\begin{equation}\label{e4a}
\begin{array}{l}
k=\bigg \langle \frac{\sum (Hr_i+\delta Vs_i) r_i(1-a p_i)^{-1}}{H\sum r_i^2(1-a p_i)^{-2}}\bigg \rangle\\ 
\approx \frac{1+a^2+3a^4+15a^6+\ldots}{1+3a^2+15a^4+105a^6+\ldots}\leq 1.
\end{array}
\end{equation}   
The numerator and the denominator contain the sum of two divergent series (\ref{e3b}) and there are problems when applying the formula 
(\ref{e40}). This fact does not prevent us from using (\ref{e40}) to process mock samples.

This result is nothing new \cite{buo}. It is known that when using LSM for fitting by linear regression, errors in the velocities, i.e. 
ordinate of data points, lead to a scatter of the obtained values of the slope angle, but not to a bias. On the contrary, errors in 
distance estimate, i.e. the abscissa of data points, lead to a systematic underestimation of this angle.
In the considered case, the underestimation is determined only by the parameter $a$. For $a = 0.2$ we have an estimate for the mean 
value of the coefficient $k = 0.96$, and for $a = 0.3$ we have $k = 0.92$. Note that this bias does not disappear at large sample size.

\subsection{Demonstration of the effect using Monte Carlo simulations. Data and routine}\label{ss23}

It seems to me that direct demonstration of the effect is more convincing than theoretical estimates, especially if latter are not 
particularly simple.
So I use the Monte Carlo (MC) method to demonstrate this effect and to clarify some important details. 

I process 1000 mock samples. 
I do not use any real observational data, but I take ranges of distances close to the real sample for galaxies from the 
Flat Galaxies Catalogue (FGC) \cite{FGC} and the Revised Flat Galaxies Catalogue (RFGC) \cite{RFGC}. I provide the results for 
the values of the parameters close to those obtained 
when processing real data, but I carry out calculations also for different parameter values, varying them within reasonable limits. 
In all cases the effects remain the same qualitatively. 

Mock samples are generated as follows. The values $v_i$ were determined
according to formula (\ref{e1}), corresponding to a preselected set of distances $r_i$. Then 1000 random sets of $V_i$ and $R_i$ were 
obtained from them, calculated by equations (\ref{e2}) and (\ref{e3}) or (\ref{e3a}). I used the values $a = 0.2$ and 
$\delta V = 1000$ km s$^{-1}$ as a main choise. For details on how these parameters were obtained, see \cite{2013Ap&SS.343..747P}. The values of the coefficient $k$ 
and its root-mean-square error $\Delta k$ are determined using the LSM formulae for each of these mock samples. In all cases, the 
average value of $k$ is 0.962 in accordance with (\ref{e4}).

To show the results obtained I use a dataset of 1402 galaxies located at a distance of $30h^{-1}$ Mpc to $100h^{-1}$ Mpc. Two 
galaxies are located at each distance interval of $0.1h^{-1}$ Mpc. Their Hubble velocities, calculated by (\ref{e1}), lie in the range 
from 3000 km s$^{-1}$ to 10000 km s$^{-1}$. Naturally, such a set has nothing to do with the real distribution of distances 
to galaxies. The impact of this distribution is discussed in Sec. \ref{ss26}. In the meantime, we are talking only about demonstrating 
the effect.

\subsection{Demonstration of the effect using Monte Carlo simulations. Results}\label{ss23a}

\begin{figure}[tb]
\includegraphics[width=\columnwidth]{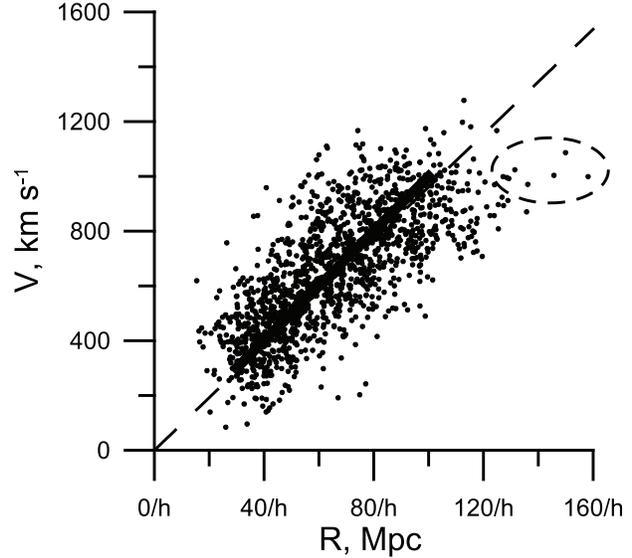}
\caption{$V_i$ and $R_i$ data for one of the mock samples. The initial points $v_i(r_i)$ lie on the thick straight line. 
The straight dashed line drawn by the LSM passing through the origin}
\label{f1}
\end{figure}

Fig. \ref{f1} shows one of the mock samples $V_i(R_i)$ obtained by adding noise to this original dataset. It also shows a thick straight 
line on which the initial points $v_i(r_i)$ lie and a straight dashed line drawn by the least squares method passing through the origin. 
The reason for underestimating the coefficient due to the abscissa changing is, in particular, the circled group of points on the right 
edge of the graph. In Fig. \ref{f1} we see a specific group of points for one of a mock sample. But for almost any mock sample, 
there is a similar group lying to the right of the straight line describing the Hubble law, which can be confused with outliers. 
The reason for this is discussed in Sec. \ref{ss24} and shown in Fig. \ref{f3}.

\begin{figure}[tb]
\includegraphics[width=\columnwidth]{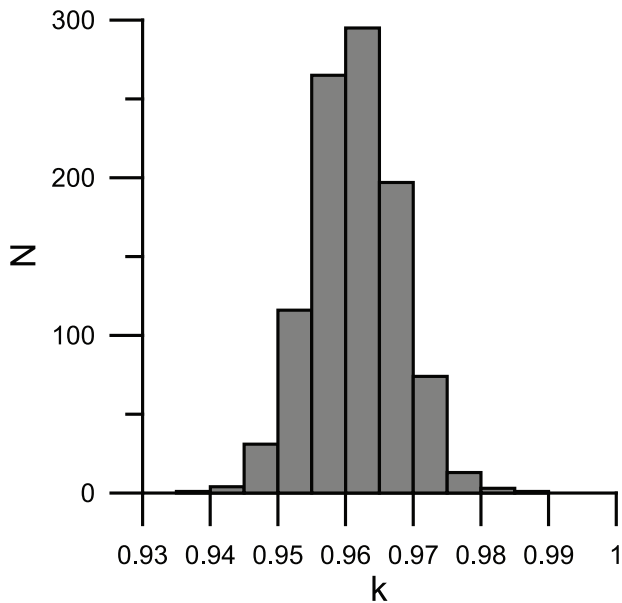}
\caption{Histogram distribution of the $k$ parameter values for 1000 mock samples}
\label{f2}
\end{figure}

The slope was determined by formula (\ref{e40}), from which the $k$ values for each mock sample were determined.
The distribution of $k$ values over different mock samples is close to a normal one. It is shown in Fig. \ref{f2}. They was less than 1 
for all 1000 mock samples and the mean value of the Hubble constant was underestimated by about $4 \%$. The values of the
LSM errors of $k$ are obtained in all cases using the standard LSM formula. Their typical value is  
$\Delta k\approx 0.1\%$. \textit{Thus, when processing any of mock samples, the underestimation of the value of the 
Hubble constant significantly exceeded the nominal error of this value, obtained using the standard data processing procedure.}

Smaller average values of $k$ were obtained when using noise of the form (\ref{e3a}). For $a = 0.1$ MC simulations provide the value 
$\langle k \rangle = 0.98$, for $a = 0.15$ $\langle k \rangle = 0.956$, for $a = 0.2$ $\langle k \rangle = 0.91$, for $a = 0.25$ 
$\langle k \rangle = 0.80$, and for $a = 0.3$ $\langle k \rangle = 0.58$. In this case, the impact of the effect under consideration 
increased.

Is it possible to obtain a more adequate result when processing, having an estimate of the values $a$ and $\delta V$? Let us try 
the maximum likelihood estimation (MLE). The formulae for calculating the slope of a straight line passing
through the origin and fitting the $V_i,R_i$ sample are derived in Sec. \ref{ss22a}. 
The MLE at reasonable values of $a$ gives results even more deviating from the true ones than those obtained from the LSM. Using 
the condition (\ref{e7a}) for processing the mock sample, shown in Fig. \ref{f1}, one can get $k=0.93$. This sample was generated using
(\ref{e2}) and (\ref{e3}) with $a=0.2$. The equation (\ref{e7b}) provides the value $k=0.88$ for the mock sample prepared using (\ref{e2}) and (\ref{e3a}) with $a=0.2$.
So, MLE has no advantages over LSM in this case.

Therefore, in what follows, I will use only LSM for data processing and add noise in accordance with (\ref{e3}). More 
sophisticated methods of parameter determination and more accurate treatments of measurement error and noise sources
will only complicate the presentation. Qualitative conclusions about effects are independent 
of these details.

\section{Overestimation on the Hubble constant for subsamples with galaxy distances limitation}\label{s3}

\subsection{An impact of data selection when limiting the range of the distance indicator}\label{ss24}

Astronomers processing real observational data can get some $V(R)$ dependence similar to that shown in Fig. \ref{f1}. They may have 
a rather natural idea to process not the entire sample, but its subsample obtained by limiting the range of $R$ variation from 
above or below, i.e. subsample with $R>R_{min}$, $R<R_{max}$ or $ R_{min}<R<R_{max}$. And they have many reasons to do so.
 
At small distances the Hubble velocities do not exceed the velocities of random peculiar motions; it makes sense to exclude the 
influence of the Local Group, not to mention the natural limitation $R>0$, which can be violated, although very rarely, by noise 
of the type (\ref{e3}) at large random deviations $p_i$. At the far end of the sample there is an influence of the incompleteness of 
the sample or its asymmetry, in particular, due to the difference in observations in the two celestial hemispheres.
                                                               
An astronomer dealing with the sample shown in Fig. \ref{f1} may decide to discard the data corresponding to the group of points 
surrounded by a dotted oval. One of the possible options is to process a subsample with $R_{max} = 120 h^{-1}$ Mpc.

However, by limiting the range of variation of the distance indicator $R$, one introduces some data selection into the subsamples. 
This leads to a statistical effect similar to the well-known Malmquist bias \cite{1922MeLuF.100....1M,1925MeLuF.106....1M}.The influence of selection effects related 
to this has been examined in \cite{2008AN....329..864P}. What is the mechanism of this selection? It is not difficult to illustrate it.
Consider a subsample with upper-bounded distance indicators $R_i<R_{max}$ and a subsample in which a similar constraint is applied 
to the true distances $r_i<R_{max}$. In the latter, the slope of the dependence $v(r)$ is by definition equal to the Hubble parameter; 
when passing to the dependence $V(R)$ it is determined from (\ref{e40}). 

\begin{figure}[tb]
\includegraphics[width=\columnwidth]{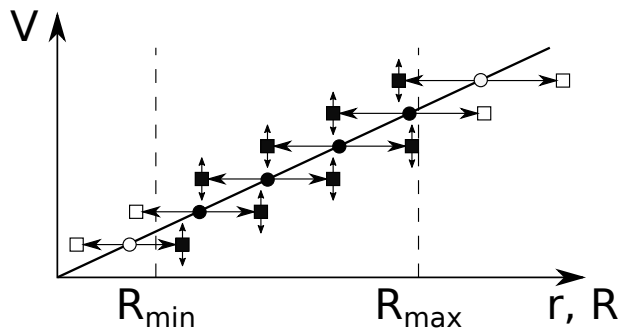}
\caption{Explanation of the data selection mechanism caused by cropping the range of variation of the distance indicator to galaxies}
\label{f3}
\end{figure}

The line in Fig. \ref{f3} corresponds to the Hubble law $v = Hr$. It contains the sample points $v_i, r_i$, indicated by circles.
Distance limits are shown with vertical dashed lines. 
The sample $V_i, R_i$ is plotted on the same graph. Adding noise (\ref{e3}) shifts circles to the right or left, and noise (\ref{e2}) 
up or down. This corresponds to the arrows in Fig. \ref{f3}. The squares show the positions of points with coordinates 
$v_i, R_i$, i.e. an averaged position of the sample points $V_i, R_i$. 

If we consider two subsamples bounded by two vertical lines 
and fill in black the symbols of the points that fall there, then black circles will fall into the subsample with $R_{min}<r<R_{max}$, 
and the subsample with $R_{min}<R<R_{max}$ will get values shifted randomly up or down relative to the black squares. This is effect of 
non-Hubble motions. White circles 
and squares will be excluded from the corresponding subsamples. From the Fig. \ref{f3} it is easy to see how the centres of the possible 
location of the sample points $V_i,R_i$ are located relative to the straight line.

Thus, the subsample with $R_i<R_{max}$ includes some black squares obtained by shifting white circles that are missing in the 
$r_i<R_{max}$ subsample, these are points with $R_i<R_{max}<r_i$. When adding noise according to formula (\ref{e3}), they shift to 
the left in Fig. \ref{f3}. A noise like (\ref{e2}) shifts their position up or down, but on average they are above the straight 
line representing the Hubble law (\ref{e1}). In addition, some black circles that are present in the $r_i<R_{max}$ subset are 
discarded. They are shifted to the right so that $r_i<R_{max}<R_i$ and, on average, would be below the straight line. 

So, in comparison with the sample $r_i<R_{max}$, the sample $R_i<R_{max}$ has additional points above the line $v = Hr$ and a 
number of points below it are excluded from the subsample. Therefore, the slope of the straight line approximating the points of 
this subsample is increased in comparison with the slope of the full sample (\ref{e40}). This is the effect of selection and it is 
caused by the statistical nature of distance indicators described by formula (\ref{e3}). 

It is easy to show from similar 
reasoning that setting the lower limit of the distance indicators $R$ also leads to some selection. However, in this case the 
additional points are located mainly under the line and the discarded points are above it. In addition, in accordance with 
(\ref{e3}), the horizontal shifts near the lower boundary are significantly less than near the upper boundary; this weakens the 
influence of selection due to the establishment of the lower boundary.
\begin{figure}[tb]
\includegraphics[width=\columnwidth]{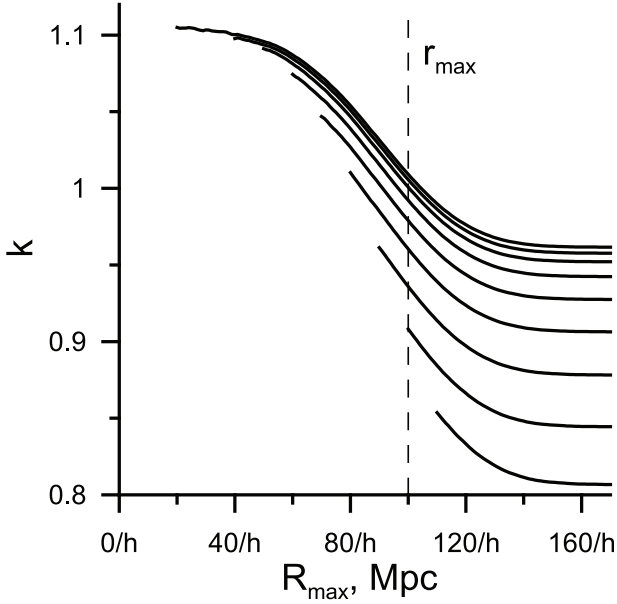}
\caption{Dependence of $\langle k \rangle$ on $R_{max}$ at fixed $R_{min}$. The lines (from top to 
bottom) correspond to $R_{min}= 20,\; 30,\; 40,\; 50,\; 60,\; 70,\; 80,\; 90$ and $100 h ^{-1}$ Mpc}
\label{f4}
\end{figure}
\begin{figure}[tb]
\includegraphics[width=\columnwidth]{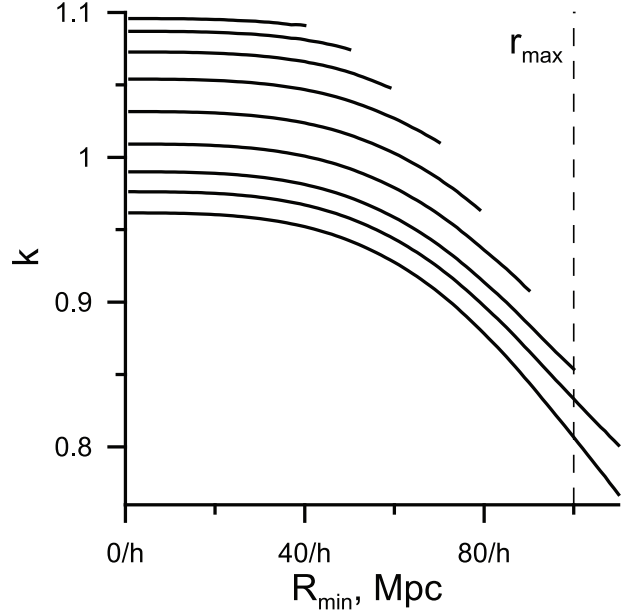}
\caption{Dependence of $\langle k \rangle$ on $R_{min}$ at fixed $R_{max}$. The lines (from bottom to top) 
correspond to $R_{max}= \infty,\; 120,\;  110,\; 100,\; 90,\; 80,\; 70,\; 60$ and $50 h ^{-1}$ Mpc}
\label{f4a}
\end{figure}

\subsection{Monte Carlo simulations. Data and routine}\label{ss24a}

I use the Monte Carlo simulations to produce a lot of $R_i$ and $V_i$ sets and then cut off all data outside the preselected limits 
of distance indicator to obtain mock subsamples. After processing them by LSM I get a set of $k$ coefficients according to (\ref{e40})
and calculate their mean value. This value is determined not only by the original sample, but also by the boundaries of the subsample.

What is it for? Each set of random deviations $p_i, s_i$ gives a random mock sample, after cutting from above and below we get a set 
of mock subsamples. The values $k$ calculated for these subsamples are described by a distribution close to normal. It is characterised 
by its mean value $\langle k \rangle$ and the standard deviation $\sigma$. According to the central limit theorem, the mean value of $k$ 
obtained from $N$ subsamples is $\langle k \rangle$, and its root-mean-square deviation from the mean is close to $\sigma N ^{-1/2}$. 
The error calculated by the LSM drops to zero with an increase in $N$, and can become significantly less than the deviation from the 
true value $k=1$. \textit{This could lead to a significant overestimation of the accuracy of the value obtained by the LSM, in particular, 
of the Hubble constant.}

Naturally, we are primarily interested in the value $\langle k \rangle$. It depends on a number of parameters, primarily $a$. Obviously, 
for $a \to 0$ we have $\langle k \rangle \to 1$. The deviation from $\langle k \rangle = 1$ monotonically increases with the growth 
of the parameter $a$. The value of $\langle k \rangle$ slightly depends on the distribution of the initial distances $r_i$. We will 
discuss this issue a little further when we consider the distribution of distances to galaxies in the realistic sample and the effect 
of its incompleteness.

In addition, the value of $\langle k \rangle$ is affected by details of generating the subsample such as the presence or absence of 
clipping at the top and bottom and the position of the boundaries of these clipping. I use the Monte Carlo method to study this 
dependence. It is much simpler and intuitive than complex analytical calculations.

The simplest initial sample is used: 991 galaxies with distances forming an arithmetic progression with an interval of 
$0.1 h^{- 1}$ Mpc. The distance to the most distant galaxy is $r_{max}= 100 h^{-1}$ Mpc, to the nearest one is 
$r_{min}=0.01 r_{max}= 1 h^{-1}$ Mpc. In Figs. \ref{f4} and \ref{f4a} the lower border lies slightly to the right of the ordinate 
axis, and the upper one is indicated on it by a vertical dashed line.

I have added random variations to the initial data using (\ref{e1}-\ref{e3}) with parameters $a = 0.2$ and 
$\delta V = 1000$ km s$^{-1}$. The subsamples are formed by simple cropping the 
same 1000 mock samples, which excludes the possibility of influencing the result by including or excluding individual points.

\subsection{Monte Carlo simulations. Results}\label{ss276}

Figs. \ref{f4} and \ref{f4a} show the $\langle k \rangle$ dependence on $R_{max}$ and $R_{min}$ values.
Before discussing its details I want to emphasize that all points on 
them are obtained by processing different subsamples of the same mock samples using LSM. 
Nevertheless, the average $k$ values differ by ten percent or more. So, processing details are important.

Fig. \ref{f4} shows the mean $\langle k \rangle$ versus $R_{max}$ value for subsamples with different fixed values of $R_{min}$. 
The lines correspond to the different $R_{min}$ values. Top one corresponds to $R_{min}= 20 h^{-1}$ Mpc and almost coincides with the line 
drawn for subsample without cut-off from the bottom. Then there are lines with $R_{min}= 30 h ^{-1}$ Mpc, 
$40 h^{-1}$ Mpc, 50 $h^{-1}$ Mpc, $60 h^{-1}$ Mpc, $70 h^{-1}$ Mpc, $80 h^{-1}$ Mpc, 
$90 h^{-1}$ Mpc, and $100 h^{-1}$ Mpc. 
I discard the leftmost edges of the curves, corresponding to subsamples with a small number of points by reason of a slight difference 
between $R_{max}$ and $R_{min}$. 

Right parts of curves reach plateaus. This is because of the effect of selection associated with 
the upper bound of the subsample disappears at $R_{max}\gg r_{max}$. The height of the plateau depends on $R_{min}$, increasing 
as it decreases. At $R_{min}\ll r_{min}$, the influence of the selection associated with the lower cutoff limit disappears. 
In the absence of selection associated with both cuts, we get $\langle k \rangle=0.96$ in full accordance with (\ref{e4}). This 
corresponds to the height of the upper plateau. From Fig. \ref{f4} it can be seen that the value $\langle k \rangle$, obtained by 
the LSM, decreases monotonically with increasing either $R_{max}$ or $R_{min}$. This dependence is associated with an explanation 
of the reason for the selection effect and follows from Fig \ref{f3}.

The left edges of the curves correspond to values at which the subsample still contains the minimum reasonable number of data points.

For clarity, Fig. \ref{f4a} shows $\langle k \rangle$ as a function of $R_{min}$ for subsamples with different fixed $R_{max}$. 
Lines from bottom to top correspond to the values $R_{max}=\infty, 120, 110,100,90,80,70,60,50\; h^{- 1}$ Mpc.
 
Let me remind you that the initial points before adding noise corresponded to the distances to galaxies from 
$r_{min}= 30 h ^{-1}$ Mpc to $r_{max}= 100 h ^{-1}$ Mpc. The lower limit lies to the left of the curves in 
Fig. \ref{f4} and the upper one is indicated on it by a vertical dashed line. The number of points to the left of $r_{min}$ is 
insignificant, so it depends little on $R_ {min}<r_{min}$. Therefore, in Fig. 4 lines corresponding to $R_{min}= 20 h ^{-1}$ Mpc
and no clipping almost coincide and are represented by one upper curve. The $\langle k \rangle$ value decreases with increasing 
$R_ {min}$ and constant $R_{max} $. This is caused by two reasons: a decrease in the number of points far from the edges and an increase 
in abscissa errors near the lower cutoff edge due to (\ref{e3}).

As $R_{max}$ increases, the number of points increases and the average value increases both for this reason and due to the influence of 
the above-described effect caused by the upper bound. At $R_{max} \gg r_{max}$, the number of points included in the subsample 
decreases greatly with a further increase in $R_{max}$ and the curve reaches a plateau. Its value depends on $ R_{min}$ and at 
from $R_{min} \ll r_{min}$ tends to the value (\ref{e4}).

\begin{figure}[tb]
\includegraphics[width=\columnwidth]{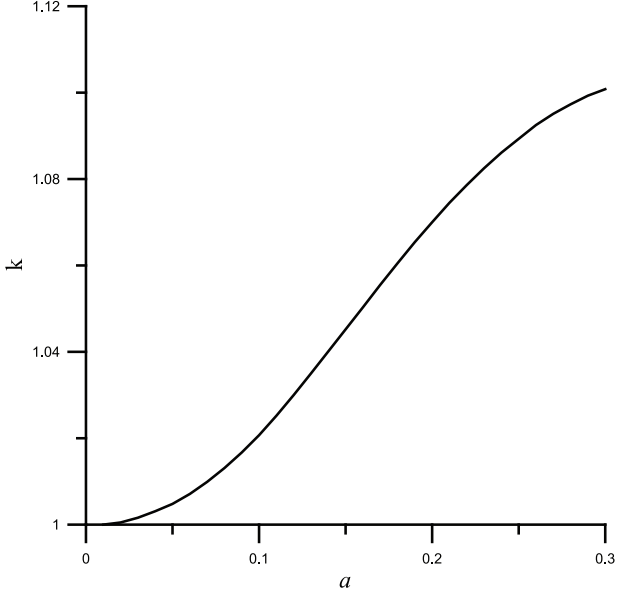}
\caption{Dependence of $\langle k \rangle$ on $a$ for the subsample with $R_{min} = 30 h^{-1}$ Mpc  
$R_{max} = 70 h^{-1}$ Mpc }
\label{f5}
\end{figure}

Figs. \ref{f4} and \ref{f5} show the results of LSM processing of the subsamples obtained by cutting $R_{max}$ and $R_{min}$. They also include 
some exotic variants with a strange choice of these values, which are unlikely to be used in real astronomical data processing. 
I want to return to the question of dependence of $\langle k \rangle$ on $a$ and consider a subsample with reasonable boundaries 
$R_{min} = 30 h^{-1}$ Mpc, $R_{max} = 70 h^{-1}$ Mpc. The value $\delta V= 1000$ km s$^{-1}$ does not change, 
the initial $r_i$ set and other details are the same as for the previous calculations in this section. The plot of 
$\langle k \rangle$ versus $a$ is shown in Fig. \ref{f5}. It can be seen from it that $k$ is overestimated by 2\% at $a = 0.1$ and by 
5\% at $a = 0.16$. Thus, the introduction of the upper or/and lower bound significantly changes 
the obtained mean value of $k$. 

\begin{figure}[tb]
\includegraphics[width=\columnwidth]{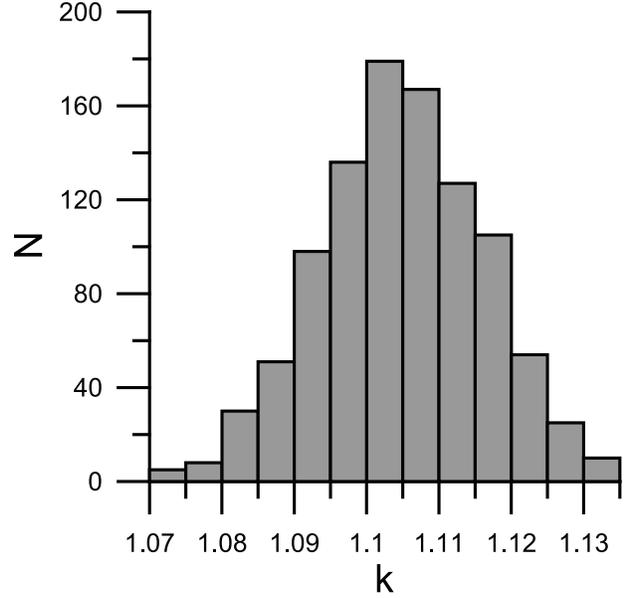}
\caption{Histogram distribution of the $k$ parameter values for 1000 subsamples obtained from mock samples by distance limitation}
\label{f6}
\end{figure}

In this case the distribution of $k$ is also close to normal as one can see in Fig. \ref{f6}. Note that the histogram is based on the 
results of simulations with the same noise parameters, but for a different initial dataset. It contains more than 21400 points 
modeling galaxies, evenly distributed from $r_{min}=10 h^{-1}$ Mpc to $r_{max}> 1080 h^{-1}$ Mpc. A subsample 
is chosen that is sufficiently distant from these boundary values, having $R_{min}=50 h^{-1}$ Mpc $= 5 r_{min}$ and 
$R_{max} = 90 h^{-1}$ Mpc$ <0.09 r_{max}$ and containing over 800 mock data. The exact number for each simulation is 
determined by a set of random variables $p_i$ and ranges from 776 to 892 for 1000 mock samples used to construct Fig. \ref{f6}. 
The scatter of values for different subsamples is greater than in Fig. \ref{f2}. However, they are all not only greater than 1, but 
even greater than 1.07. The obtained distribution of values can be considered typical for subsamples with the lower and upper boundaries 
removed from the boundaries of the original distribution by distances. 

\begin{figure}[tb]
\includegraphics[width=\columnwidth]{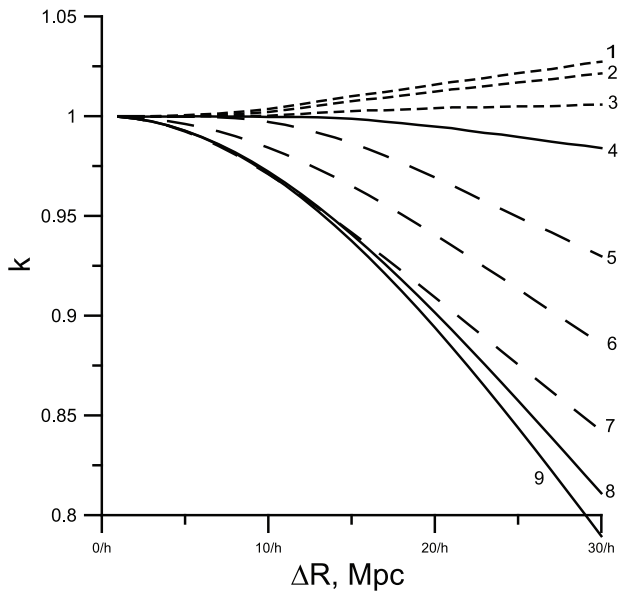}
\caption{Dependence of $\langle k \rangle$ on $\Delta R$ with errors of distances in the form (\ref{e9}) for subsamples with 
different $R_{min}$ and $R_{max}$. The numbered curves 1-9 correspond to the following $R_{min}$ values: 0, 10, 20, 30, 30,
30, 30, 0 $h^{-1}$ Mpc and without cropping from the bottom. The $R_{max}$ values are 70, 70, 70, 70, 85, 100, 120
$h^{-1}$ Mpc and two curves without cropping from the top}
\label{f7}
\end{figure}

\textit{Monte Carlo calculations confirm that selection caused by the upper limit $R_{max}$ leads to overestimating of $k$, while selection 
associated with the lower limit $R_{min}$ to it underestimating.} The first effect is superior to the second with a reasonable choice 
of cutoff boundaries for errors of the form (\ref{e3}).
The impact of data selection because of an upper limit on the distance indicator to the galaxy is several times greater than the 
effect previously described in Secs. \ref{ss22} - \ref{ss23a}. Note that the source of both is using of statistical relations 
to obtain a galaxy distance estimate independent of the redshift. The standard deviation of the mean for a sample of $N$ objects 
decreases with increasing $N$ in proportion to $N^{-1/2}$ and can become very small. But this average is biased and one deals with 
a classic example of an estimation with high precision and low accuracy.

\subsection{Influence of distance distribution of galaxies}\label{ss26}

In the previous simulations I used the initial uniform distribution of galaxies over a certain interval of distances. This is not 
valid for real samples. At small distances the number of galaxies falling within the range of distances from $r$ to $r + dr$ is 
proportional to $r^2$ simply because they fall into the volume $4\pi r^2 dr$. A number of far galaxies begins to decrease with 
distances due to the incompleteness of the sample, which simply does not include very distant galaxies. Naturally, a similar 
dependence is observed for the distribution of galaxies over the values of the distance indicators $R$.

I want to discuss the impact of both of these effects. They not only affect the effect caused by limiting the sample from above, 
but also lead to the appearance of another type of errors associated with data selection.

If we restrict the subsample by the condition $R<R_{max}$ and it has a high degree of completeness, then, compared to the uniform 
initial distribution of galaxies over distances, a large part of galaxies are located near the upper boundary and the influence 
of the selection discussed in the section is intensified. It is easy to verify this by applying the Monte Carlo method to a sample 
of 2815 galaxies distributed with a number proportional to $r^2$. At a distance of 30 $h^{-1}$ Mpc there are 36 galaxies, at a 
distance of 35 $h^{-1}$ Mpc there are 49 more galaxies, and so on up to a distance of 100 $h^{- 1}$ Mpc, where 400 galaxies are 
located. A random noise is added to this initial sample according to (\ref{e2}) and (\ref{e3}). The resulting 1000 mock samples are 
processed using LSM formulae. The results obtained are perfectly described by formula (4), as expected. For $a = 0.2$, the average 
value is $k = 0.96$, for $a = 0.25$ $k = 0.94$ and for $a = 0.3$ on average $k = 0.92$.

But the mean values of $k$ obtained for subsamples with $R_{max}=80 h^{- 1}$ Mpc turn out to be slightly larger than with the 
initial uniform distribution. For $a = 0.15$ on average $k = 1.08$, for $a = 0.2$ $k = 1.12$, for $a = 0.25$ $k = 1.16$ and for 
$a = 0.3$ on average $k = 1.19$.

With an increase in $R_{max}$, the influence of selection effects associated with sample incompleteness begins to take effect. 
As I mentioned in the Section \ref{ss21a}, statistical dependencies allow us to estimate a certain quantity that does not depend on 
the distance to the galaxy, for example, its absolute luminosity or linear size. Then, the photometric distance or angular distance 
is obtained from it and observable quantities such as apparent magnitude or angular dimensions.

Since the dependencies are statistical, the luminosity or size of each individual galaxy deviates from the average estimate. If they 
are larger or brighter, then we underestimate the distance to the galaxy, considering it closer than in reality. And if it is weaker 
or smaller in size, then we overestimate the distance. In formula (\ref{e3}) this corresponds to negative and positive values of 
$p_i$, and in Fig. \ref{f3} to left and right shifts. 

However, a brighter or larger galaxy is more likely to get into the sample than a faint small galaxy. Therefore, it should be expected 
that the number of galaxies shifted to the right will exceed the number of galaxies shifted to the left. The distribution of the 
$p_i$ values for the galaxies included in it will still be random, but it will not only be different from Gaussian, but also have 
a nonzero mean. As a result, the sample will include fewer galaxies lying above the line $V = HR$ and more ones lying below it. 
This will lead to an additional underestimation of the value of $k$ obtained by processing the data using the LSM versus complete sample. 

As one can see, this selection acts in exactly the opposite way than the selection associated with the consideration of subsamples 
limited by distance. Note that both in deriving (\ref{e4}) and in simulations, I did not consider the influence of this effect in the 
Sect. \ref{ss24}, assuming that a galaxy can move in the $V, R$ plane relative to its true position, but not disappear from the sample. 

\section{Impact of distance-dependent and independent errors}\label{ss24b}

For completeness, I also performed similar simulations with the distance estimation errors in the form 
\begin{equation}\label{e9}
R_i=r_i+\Delta R p_i.
\end{equation}
In this case the distribution of deviations in the estimation of the distance does not depend on this distance, in contrast to the 
previously considered errors of the form (\ref{e3}) or (\ref{e4}). So, the effect associated with the difference in magnitude of 
errors for nearby and distant galaxies disappears. But this does not mean that there is no difference between the effect of cutoffs 
at the upper bound of the subsample, described by $R_{max}$, and at the lower, characterized by $R_{min}$. The reason is that we 
fit the data with a relationship in the form (\ref{e1}), i.e. a line through the origin. Naturally, the origin is nearer to a 
lower border.
 
I apply the Monte Carlo method to the initial dataset with data for 991 mock galaxies, which is similar to the one used in the 
previous section and apply noise to it 1000 times. I add random errors of the form (\ref{e2}) with the same value $\delta V = 1000 $ km s$^{-1}$ 
and the form (\ref{e9}) with different values of $\Delta R$. A bias also exists in this case.
   
I do not show graphs similar to Fig. \ref{f4} or \ref{f4a}, and go straight to the analogue of Fig. \ref{f5}, i.e. dependencies of 
$\langle k \rangle$ on $\Delta R$ for various cutting of a noisy sample. In this Fig. \ref{f7}, solid lines depict three natural 
choices of the values $R_{max}$ and $R_{min}$. These curves are indicated by the numbers 9, 8 and 4. Curve 9 corresponds to the 
absence of any clipping. For it, from (\ref{e40}) by a method similar to (\ref{e4}), it is easy to obtain a theoretical estimate
\begin{equation}\label{e400}
\begin{array}{l}
k=\frac{\langle A\rangle}{H}= \bigg \langle \frac{\sum (Hr_i+\delta Vs_i)(r_i+\Delta R p_i))}{H\sum (r_i+\Delta R p_i)^2} 
\bigg \rangle \\
\approx \bigg(1+\frac{\Delta R^2}{U^2}\bigg)^{-1}.
\end{array}
\end{equation}   
Here $U^2=\langle r_i^2\rangle$. For the initial sample I used for MC simulations $U\approx 60 h^{-1}$ Mpc.

However, due to errors, there could be some galaxies with distance estimates $R_i<0$. It is hard to imagine that they are not 
discarded during processing. Therefore, I also considered a subsample with $R_{min} = 0$ without cutting from the top. It is easy 
to understand that the obtained values of $k$ will be larger for it than (\ref{e400}) due to the fact that points in the region 
$R<0$ provide negative contribution to the numerator (\ref{e400}) on average. By discarding them, we increase $k$ and decrease 
its deviation from 1. This subsample corresponds to the curve 8 in Fig. \ref{f7}.

I use the bounds $R_{min} = 30 h^{-1}$ Mpc, $R_{max} = 70 h^{-1} $ Mpc, i.e. those that were used to calculate the graph in 
Fig. \ref{f5}, as the third natural choice (the curve 4 in Fig. \ref{f7}). One can see that in this case the deviation of $k$ 
from 1 is significantly less than for the full sample, for which it is well described by equation (\ref{e400}). This 
demonstrates that the influence of data selection is significant even for errors of the type (\ref{e9}). Comparing Fig. \ref{f7} 
and Fig. \ref{f5}, one can see that errors of the form (\ref{e3}) lead to an overestimation of $k$, and those of the form (\ref{e9}) 
to its underestimation at the same subsample bounds.

Some additional lines are drawn in Fig. \ref{f7} to show how the values of $R_{max}$ and $R_{min} $ affect $k$. The dashed lines 1-3 
correspond to different values of $R_{max}$ with the same value of $R_{min} = 30 h^{-1}$ Mpc and the lines 5-7 with longer dashes 
correspond to different values of $R_{min} \ge 0$ with the same value of $R_{max} = 70 h^{-1}$ Mpc. For curves 4, 3, 2, 1 
$R_{min} = 30, 20, 10, 0 \; h^{-1}$ Mpc. It can be seen that with a decrease in $R_{min}$, the average $k$ increases and underestimation 
is replaced by overestimation. Lines 4, 5, 6, 7 correspond to the upper bounds with $ R_{max} = 70, 85, 100, 120 \; h^{-1}$ Mpc, 
and the line for a subsample without upper distance limitation ($R_{max} = \infty$) absent on the chart is slightly above curve 9. 
It can be seen that an increase in $R_{max}$ leads to an increase in bias.

So, the bias is caused by a combination of three effects, namely the general underestimation (\ref{e4}) or (\ref{e400}) and the 
influence of selection when cutting off at the upper and lower boundaries of the subsample, the latter depend on $R_{max}$ and 
$R_{min}$. The ratio of them is different for errors (\ref{e4}) and (\ref{e9}). 

Let's try to compare the total bias for a reasonable 
choice of boundaries. For (\ref{e3}) with $a = 0.2$ the standard deviation of the distances, i.e. a value similar to $\Delta R$ 
in (\ref{e9}) for a sample with $R_{min} = 30 h^{-1}$ Mpc, $R_{max} = 70 h^{-1}$ Mpc varies from $6 h^{-1}$ Mpc at its lower 
boundary to $14 h^{-1}$ Mpc at its upper boundary. The bias is much less than 1\% for curve 4 and $\le 5\%$ for all curves in 
Fig. \ref{f7} for $\Delta R \le 14 h^{-1}$ Mpc. It can be seen that errors of the form (\ref{e3}) provide a larger bias and 
more often lead to an overestimation of the value of $k$ than that of the form (\ref{e9}).

This is an important detail that makes it possible to quite successfully apply the LSM in the case when the random error of the abscissa 
is constant. It is not met in the case of determining the Hubble parameter and the Hubble constant, when as the distance to galaxy 
increases, so does the error in its determination. 

\section{Bias correction}\label{sss}

Naturally, astronomers usually do not process observational data using standard software packages. They try to minimize possible bias 
by correction. They take into account factors that directly affect the measured quantities, such as extinction, or the 
aperture of a telescope. The data is corrected to reduce the influence of known physical effects, for example, the change in luminosity 
depending on the redshift or the curvature of space-time is taken into account. The influence of statistical factors is also taken 
into account, for example, a correction for Malquist bias.

The effect described in Sec. \ref{ss22} can be corrected by applying the so-called ``correction for attenuation'' which would multiple 
$A$ by $1 + a^2$, assuming that $a$ is known. However, you need to know the error distribution and all its parameters for effective 
bias-correction. In real measurements, deviations are usually caused by several factors. In the Section \ref{ss21} I gave an example 
of errors in estimating the distance from the Tully-Fisher dependence in the ``linear diameter --- the emission line-width'' version, 
where errors of the method itself, as well as errors of measurements of the angular dimensions and line width, led to different 
dependences of errors on distance.

It is difficult to take into account and compensate for the effect of cut-off of the sample and its incompleteness without knowing 
the distribution of objects in the sample. So we can correct the value of the determined quantity rather ambiguously. They can be 
affected during processing.

A well-known example of such an influence is provided by experiments on measuring the charge of an electron. R.A. Millikan was awarded 
the Nobel Prize in Physics in 1923 ``for his work on the elementary charge of electricity and on the photoelectric effect''. This is how 
this phenomenon is described in R. Feynman's autobiographical book \cite{Feynman}: ``One example: Millikan measured the charge on an 
electron by an experiment with falling oil drops, and got an answer which we now know not to be quite right. $<\cdots>$ It's interesting 
to look at the history of measurements of the charge of the electron, after Millikan. If you plot them as a function of time, you 
find that one is a little bigger than  Millikan's, and the next one's a little bit bigger than that, and the next one's a little 
bit bigger than that, until finally they settle down to a number which is higher. $<\cdots>$ When they got a number that was too high 
above Millikan's, they thought something must be wrong and they would look for and find a reason why something might be wrong. 
When they got a number closer to Millikan's value they didn't look so hard. And so they eliminated the numbers that were too far off, 
and did other things like that.''

Note that the history of measuring the charge of an electron is somewhat similar to the definition of the Hubble constant in another 
aspect, too. A group of researchers from the University of Vienna, headed by F. Ehrenhaft, obtained the value of the electron charge 
less than that of Millikan, and for a long time there was a kind of ``electron's charge tension'' in physics.

Thus, the correction can reduce, but not eliminate the influence of the discussed effects, and its application also can be a potential 
source of errors.

\section{Discussion and conclusion}\label{s4}

Two points can be drawn based on the results of this work. One is more general, the second refers to the Hubble tension. The first is 
associated with the result of data processing 
by any of the statistical methods used, starting with the LSM. The simplest simulations, which can be easily repeated by anyone show 
that there is a bias in defining simple quantities like a slope of a straight line running through the origin. It is associated with the error 
in estimating the quantity used as the abscissa. In our example these are the distances to galaxies. The result is a biased estimation 
which can be quite precise for a large sample, but not necessarily accurate. Bias in the estimation of a slope cannot be eliminated 
using more sophisticated statistical processing methods such as MLE. 

Naturally, one can try to estimate it and introduce a correction for bias into the resulting value, similar to how a correction for 
Malmquist bias is introduced in some astronomical calculations. However, this is not easy to do. 
Bias is caused both by a general underestimation of the type (\ref{e4}) or (\ref{e400}), and by the influence of selection due to 
truncation of the sample or its incompleteness. The impact of a cutoff is different for the upper and lower sample boundaries. The 
total bias depends on the magnitude and distribution of the errors and on the distribution of both data points and errors over abscissa. 
It could lead either to underestimation or overestimation of the obtained slope, as can be seen in specific examples using 
the Monte Carlo method. All this greatly complicates the calculation of the correction that would compensate the impact of the effect 
under consideration.

These effects can significantly bias the value of the Hubble parameter as the slope of the straight line $v(r)$, determined from the 
redshift and the estimated distances to galaxies. Bias is quantitatively characterized by the deviation of the parameter $k$ defined 
as the ratio of the calculated value of the Hubble parameter to the true one from $k = 1$. The influence of the abovementioned factors 
and some others, the influence of which turned out to be 
insignificant, is investigated both analytically and using the Monte Carlo method.

The value of the Hubble parameter and the Hubble constant obtained by LSM are underestimated in accordance with formulae (\ref{e4}) 
and (\ref{e400}). For typical precision distance indicators of $20 \% - 30\%$ the effect is about $4 \% - 8 \%$. The error of the 
Hubble parameter obtained by the least squares formula is much less than bias. The distribution of the factor $k$ is close to normal. 
All $H$ values obtained by LSM are underestimated in all sets of 1000 simulations each.

If the mock sample is additionally cut off from above by the condition $R_i<R_{max}$ or/and from below by the condition $R_i>R_{min}$, 
this will lead to additional bias of the value of the Hubble parameter obtained by the LSM. The reason for this effect is 
explained in the Sect. \ref{s3}. The results of MC simulations presented in Figs. \ref{f4}, \ref{f4a} and \ref{f7} show that 
the values of $k$ for subsamples obtained at different cutoffs of the same sample may differ significantly.
The impact of the sample cutoff can not only exceed the influence of the aforementioned 
underestimation, but also leads to a general overestimation of the Hubble parameter value by about $8\%$ when using distance 
indicators with an accuracy of $20 \%$. Note that the impact of the effect is highly dependent on the used model of the distance 
indicator error.

It can be assumed that the indicated effects can bias the value of the Hubble constant, determined by processing the data of real 
observations. Therefore, to check the adequacy of the processing methods applied in each specific case, it is advisable to carry out 
modeling using the Monte Carlo method by adding noise to the set of initially accurate data.

From a practical point of view, we are talking about the following sequence of actions. When processing data, one of the measured values 
is considered as a function of the rest of the measured values, considered as arguments of this function, and the set of parameters that 
we are calculating. Initially, the values and errors of all parameters are determined using conventional data processing. Then a mock 
sample is created. For this, the values of the arguments are taken from the sample used. They coincide with the measured values after 
the necessary corrections have been made, if any. The value of the quantity, considered as a function, is calculated for each set of 
these arguments using the formulae used in the processing; the parameters obtained in the first stage are applied. 

Then the Monte Carlo method is used. The characteristic values of the random deviations of the arguments are selected in accordance 
with their errors. At the final stage, the bias of the each parameter is estimated. It is determined as the difference between the 
parameters obtained at the final and first stages. If necessary, one can try to correct the values of calculated parameters taking 
into account the obtained bias estimate.

The more specific conclusion is related to the Hubble tension. Astronomers use a complex ladder of distances, many steps of which are 
based on some statistical dependence. Errors associated with them accumulate as the number of steps increases. Therefore, the values 
obtained using such distance estimates can have significant bias. In particular, estimates of the Hubble constant may have low accuracy 
in spite of high precision.

As I already mentioned, the difference between the Hubble constants for the early and modern Universe is 8\% \cite{r}. The 
overestimation of this value caused by the effects discussed in the paper could be from 8\% to 12\% for a reasonable choice of the 
boundaries of the subsample. Thus, it is quite capable of explaining Hubble tension on the observed level.

The estimates obtained in this work using the Monte Carlo method demonstrate that the effect caused by bias during data processing can 
exceed the difference in the estimates of the Hubble constant values for high-z and low-z observations. Occam's razor principle 
suggests not looking for a more complex explanation for a phenomenon that can still be explained by measurement and processing errors 
leading to a bias of the Hubble constant. Especially in comparison with alternative ones, which imply a change in the foundations of 
physics or the existence of fundamentally new entities in our Universe.

\section{Appendix}

\subsection{Some mathematical features of two types of errors in estimating distances to galaxies}\label{ss21a}

Inverse transformations from $R$ to $r$ for (\ref{e3}) and (\ref{e3a}) coincide with transformations (\ref{e3a}) and (\ref{e3}), 
respectively, with the replacement $r\leftrightarrow R$ and change of the sign of the $a$ parameter. The last detail is absolutely 
unimportant because of the symmetry of the Gaussian distribution of $p_i$. So the distribution of $R$ values obtained at a fixed value 
of $r$ by formula (\ref{e3}) completely coincides with the distribution of $r$ values obtained at a fixed value of $R$ by formula 
(\ref{e3a}). The same statement remains true after the interchange of (\ref{e3}) and (\ref{e3a}). 

However, we use the values of $r$ and $R$ in an apparently asymmetrical manner when generating mock samples. The initial distribution 
of $r$ values is chosen in advance, so that this value is fixed for each galaxy. We are interested in the distribution of the $R$ 
values generated by the formulae (\ref{e3}) and (\ref{e3a}) for a fixed $r$. Their properties are different in some details. Let me 
point out these differences. 
The distribution of $R$ values described by (\ref{e3}) is normal, described by (\ref{e3a}) is non-Gaussian and asymmetric.

Note that $R$ becomes negative for large deviations, namely for $p<-p_0$ with $p_0 =|a|^{-1}$ in (\ref{e3}) and for $p>p_0$ in 
(\ref{e3a}). In this case, this value turns to 0 for $p=-p_0$ in (\ref{e3}), and becomes infinitely large for $p=p_0$ in 
(\ref{e3a}). This happens very rarely. For $a = 0.2$, this corresponds to a deviation of $5 \sigma$ in a certain direction and occurs with 
a probability $\sim 3\cdot 10^{-7}$. It is practically not realized when preparing 1000 mock samples. In principle, such problems 
theoretically exist for any case of a normal distribution, and usually deviations at the $5\sigma$ level are discussed only to show 
that they are not random. If this small probability is realized during a mock sample generation, one just need to regenerate the sample.

It is easy to assume a distribution of errors that does not allow negative values of $R_i$ and coincides with (\ref{e3},\ref{e3a})
in the first term of the expansion in the Taylor series in $a$. This is the model with lognormal multiplicative errors
\begin{equation}\label{e3aa}
R_i=r_i\exp(ap_i).
\end{equation}   
It is obvious that all the described types of bias are also typical for it. Since we are more interested in simple quantitative estimates, 
there is no point in multiplying entities beyond necessity and considering this model additionally, because the qualitative conclusions 
have to remain the same when using it.

However, even this theoretical possibility of generating very large $R$ values leads to unpleasant consequences. All moments of the 
distribution $R$ obtained for a fixed value of $r$ become infinitely large. I show this using the fact that for a Gaussian distribution 
all odd plain central moments are equal to 0 and the mean values of $p_i^n$ are equal to $(n-1)!!$ if $n$ is even. 

So, the average estimate of the distance to galaxies become infinitely large after adding noise of the form (\ref{e3a}), as well as 
higher powers of this quantity. Really, for given $r_i$ one have
\begin{equation}\label{e3b}
\begin{array}{l}
\langle R_i\rangle =r_i \langle (1-ap_i)^{-1}\rangle =r_i \langle 1+ap_i+a^2p_i^2\\
+a^3p_i^3+a^4p_i^4+a^5p_i^5+a^6p_i^6+\ldots\rangle=r_i (1+a^2\\
+3a^4+15a^6+\ldots)\\
=r_i\sum_{n=0}^{\infty}(2n-1)!!a^{2n}=\infty,\\
\langle R_i^2\rangle =r_i^2 \langle (1-ap_i)^{-2}\rangle =r_i^2\langle 1+2ap_i+3a^2p_i^2\\
+4a^3p_i^3+5a^4p_i^4+6a^5p_i^5+7a^6p_i^6+\ldots\rangle\\
=r_i^2 (1+3a^2+15a^4+105a^6+\ldots)\\
=r_i^2\sum_{n=0}^{\infty}(2n+1)!!a^{2n}=\infty,
\end{array}
\end{equation}
etc. Here the angle brackets mean averaging over values $p_i$. Both of these series diverge. This is evident from the fact that the 
ratio of two consecutive terms is equal to $(2n-1) a^2$ and $(2n + 1) a^2$ respectively and exceeds 1 for large $n$ and nonzero $a$.

The same feature of the relationship between $R$ and $r$ is manifested in the case when we fix the value of $R$ and study the 
distribution of the values of $r$. It is needed when using the maximum likelihood 
method. The probability density distribution for (\ref{e3a}) is easy to obtain
\begin{equation}\label{ee}
\begin{array}{l}
P(p_i)=(2\pi)^{-1/2}\exp\left(-\frac{p_i^2}{2}\right)\\
\propto \exp\left(-\frac{(R_i-r_i)^2}{2a^2R_i^2}\right).
\end{array}
\end{equation}   
It vanishes at $r\to\infty$ as it should. For relation (\ref{e3}) we get
\begin{equation}\label{ee1}
P(p_i)\propto\exp\left(-\frac{(R_i-r_i)^2}{2a^2r_i^2}\right)\xrightarrow [r_i \to \infty]{}
\exp\left(-\frac{1}{2a^2}\right).
\end{equation}
For the normal distribution of the random variable $p_i$ the probability density of the distribution $r_i$ becomes constant 
as $r_i \to \infty$, which excludes the possibility of its normalization. 

\subsection{Maximum likelihood data processing}\label{ss22a}

The correct use of MLE is impossible for the relation (\ref{e3}) because of (\ref{ee1}) and we have no choice but to use (\ref{e3a}). 
If a galaxy is located at a distance $\xi$ and has a Hubble velocity $\eta$, then the probability of its observation with a velocity 
$V_i$ and an estimate of the distance $R_i$ according to (\ref{e2}), (\ref{e3a}) is equal to 
\begin{equation}\label{e5}
\begin{array}{l}
w(R_i,V_i,\xi,\eta)=\frac{1}{2\pi\delta V a R_i}\exp\left(-\frac{s_i^2+p_i^2}{2}\right)\\
\propto \exp\left(-\frac{(V_i-\eta)^2}{2\delta V^2}\right) \exp\left(-\frac{(R_i-\xi)^2}{2a^2R_i^2}\right).
\end{array}
\end{equation}   
Therefore, the probability of observing a galaxy with parameters $V_i$ and $R_i$, assuming that the true distance $\xi$ and the 
Hubble velocity $\eta$ are related by $\eta=A\xi$, is proportional to
\begin{equation}\label{e6}
\begin{array}{l}
P_i(R_i,V_i,A)=\int_0^{\infty}w(R_i,V_i,\xi,A\xi)d\xi\\
\propto\int_0^{\infty}\exp\left(-\frac{(V_i-A\xi)^2}{2\delta V^2}\right) 
\exp\left(-\frac{(R_i-\xi)^2}{2a^2R_i^2}\right) d\xi.
\end{array}
\end{equation} 
The integrand decreases rapidly with distance from the sampling points, so this value is practically independent of the integration 
limits if they are far from the sampling points. So we can put $-\infty $ as the lower limit of integration.

The total probability for a given sample of $N$ galaxies is
\begin{equation}\label{e7}
\begin{array}{l}
P(A)= \prod_{i=1}^N P_i(R_i,V_i,A)\\
\propto\prod_{i=1}^N \int_{-\infty}^{\infty}\exp\left(-\frac{(V_i-A\xi)^2}{2\delta V^2}-\frac{(R_i-\xi)^2}{2a^2R_i^2}\right) d\xi.
\end{array} 
\end{equation} 
According to the MLE the optimal slope of the straight line passing through the origin and fitting dataset will correspond to the 
value $A = kH$ at which the value (\ref{e7}) is maximum. The corresponding complex nonlinear equation for this quantity is easily 
obtained after equating to zero the derivative of $\ln(P)$ with respect to the parameter $A$:
\begin{equation}\label{e7a}
\begin{array}{l}
\sum_{i=1}^N \frac{\int_{-\infty}^{\infty}(V_i-A\xi)\xi F(\xi,A,V_i,R_i)d\xi}{\int_{-\infty}^{\infty}F(\xi,A,V_i,R_i)d\xi}=0,\\
F(\xi,A,V_i,R_i)=\exp\left(-\frac{(V_i-A\xi)^2}{2\delta V^2}-\frac{(R_i-\xi)^2}{2a^2R_i^2}\right). 
\end{array} 
\end{equation} 

For small values of $a \ll 1$ one can estimate the integrals (\ref{e6}) using the steepest descent method and obtain
\begin{equation}\label{e8}
P_i(R_i,V_i,A)\approx (2\pi)^{1/2}aR_i\exp\left(-\frac{(V_i-AR_i)^2}{2\delta V^2}\right) 
\end{equation} 
and MLE gives the standard LSM formula. By expanding the integrand from (\ref{e6}) into a series in powers of $\xi-R_i$, one can 
obtain corrections to this expression. However, this makes no sense, since for small $a\ll 1$ it is easier to use estimate (\ref{e4}), 
and for arbitrary $a$ the value of $A$ is easier to find numerically. In any case, it is necessary to have an estimate of the accuracy 
of the distance indicators $a$ and the average peculiar velocity of galaxies $\delta V$ to obtain the Hubble constant.

One can try the following trick: use the MLE with the relation (\ref{e3}), not paying attention to the value of the normalization 
constant, which disappear after $\ln(P)$ differentiation. This gives the condition
\begin{equation}\label{e7b}
\begin{array}{l}
\sum_{i=1}^N \frac{\int_{-\infty}^{\infty}(V_i-A\xi) F_1(\xi,A,V_i,R_i)d\xi}{\int_{-\infty}^{\infty}\xi^{-1}F_1(\xi,A,V_i,R_i)d\xi}=0,\\
F_1(\xi,A,V_i,R_i)=\exp\left(-\frac{(V_i-A\xi)^2}{2\delta V^2}-\frac{(R_i-\xi)^2}{2a^2\xi^2}\right). 
\end{array} 
\end{equation} 
Note that such incorrect techniques are often used in various fields of physics and sometimes make it possible to obtain correct results.
As an example, I mention the methods of working with divergent integrals used in field theory, including the renormalization.
However, in our case, this does not give any improvement in the estimates, as can be seen from the results of the Monte Carlo simulation.

\subsection{An impact of the large-scale collective motion of galaxies}\label{ss25} 

In the previous model I neglect the effect of the large-scale collective motion of galaxies. It is not only well known, but also 
makes a significant contribution to the velocities of peculiar motions of individual galaxies. The radial component of its velocity 
depends both on the distance to the galaxy and on the direction towards it. The form of this dependence can be quite complex and 
include many terms, the values and statistical significances of which are determined when processing observational data as it is 
done in papers \cite{2001AstL...27..765P,2013Ap&SS.343..747P}. By the way, the influence of errors in determining the distance on 
the parameters of the velocity field of collective motion was investigated in \cite{2008AN....329..864P}. Knowing this field, 
it is possible to determine the density distribution of matter, 
including dark matter, in an area with a radius of about $75 h^{-1}$ around us \cite{2006AstL...32..287S}.

In this article we are interested in the effect of errors in estimating the distance to galaxies on the value of the Hubble parameter. 
How it might be affected by the large-scale collective motion of galaxies? Look at the Fig. 1. The best-fitting line not crossing the 
origin for this data 
is a line with a non-zero intercept and $k\approx 0.69$ because of an impact of errors in distance estimations. But I naturally 
approximate them with a line passing through the origin in accordance with Hubble's law (\ref{e1}). If we add to (\ref{e3}) the 
radial component of the collective galaxy motion, then it could play the role of an effective intercept, especially for a strongly 
asymmetric sample.

Let us check this hypothesis using the example of the simplest model of collective motion, in which all galaxies move as a whole at 
a constant speed $\Delta V$. For such bulk motion we have
\begin{equation}\label{e10}
v_i=Hr_i+\Delta V\cos \theta_i,
\end{equation}   
where $\theta_i$ is the angle between the apex of motion and the direction to the $i$-th galaxy. The values of the angles were chosen 
randomly, the values of $\Delta V = 200$ km s$^{-1}$ and $a = 0.2$ were used.

Three components of the collective motion velocity vector are usually determined when investigating a bulk motion. But in this toy model 
we define only one component of it towards the chosen direction. But even such a simple model helps to find out whether the collective 
motion of galaxies influences the value of the Hubble constant obtained by processing data on their velocities.

One thousand mock samples were prepared with the same errors (\ref{e2}) and (\ref{e3}) as in the previous case, but using (\ref{e10}) 
instead of (\ref{e1}). For each of them the values and errors of $H$ and $\Delta V$ were obtained from LSM. 
We are primarily interested in the Hubble constant value. It remained at the level of $95\%$ -- $97\%$ of the original. It is seen that 
the complication of the model of galaxy motion did not affect the effect.

Now let's introduce some anisotropy into the spatial distribution of galaxies. To do this, I shifted the values of $\cos \theta$ by 
adding 0.01 to each value. If the obtained value of $\cos \theta$ exceeds 1, I determine that $\cos \theta=1$. This does not affect 
the obtained value of Hubble constant. Even after shifting these values by 0.1, i.e. significantly, it remains the same. Thus, the 
collective motion of galaxies, neither by itself nor together with the sample anisotropy, affects the effect under consideration.

\vskip3mm \textit{This work was supported by the National Research Foundation of Ukraine under Project No. 2020.02/0073. 
I am grateful to Dr. V.Zhdanov for valuable advice and discussions and to reviewers unknown to me for valuable comments.}

\end{document}